\documentclass{article}
\usepackage[english]{babel}
\usepackage{eurosym}

\usepackage[usenames]{color}

\begin{document}

\title{Informatics Perspectives on Decision Taking:\\
{\large a Case Study on Resolving Process Product Ambiguity}%
}
\author{%
J. A. Bergstra\\
{\small  Section Theory of Computer Science,
Informatics Institute,} \\
{\small Faculty of Science,
University of Amsterdam, The Netherlands.}%
\thanks{Author's email address: {\tt j.a.bergstra@uva.nl}. 
This paper is a revised version of~\cite{Bergstra2011b}. A subtitle has been added which indicates the 
experimental flavor of the work. The title might have been adapted by changing ``Informatics Perspectives'' into ``Informaticological Perspectives'' (for a justification of the term informaticology
see~\cite{Bergstra2012a}), which I now consider to be more adequate than the actual title, but I have preferred not to change the title for reasons of consistency. 
Many minor improvements have been applied, 
and a terminology for choice making has been included, moreover the concluding remarks were updated.
The first version of the paper has been written  in the context of the NWO Jaquard project 
Symbiosis which focuses on software asset outsourcing and on IT outsourcing in general. The objective to
analyze outsourcing decisions (see~\cite{Delen2007}) and follow-up outsourcing decisions
(see~\cite{BLA2009}) requires a preparatory investigation of decision taking in general. The work on decision taking
will be more general than the intended application to IT sourcing issues requires it to be.}
}

\maketitle

\begin{abstract}
\noindent A decision is an act or event of decision taking. Decision making always includes decision taking, the latter not involving significant exchanges with non-deciding agents.  A decision outcome is a piece of storable information constituting the result of a decision. Decision outcomes are typed, for instance: plan, command, assertion, or boolean reply to a question. Decision outcomes are seen by an audience and autonomous actions from the audience is supposed to realize the putting into effect of a decision outcome, thus leading to so-called decision effects.

Decision outcomes are supposedly expected by the decider. 
 Using  a model or a theory concerning the causal chain leading from a decision outcome to  one or more decision effects may support a decision taker decision taker in predicting plausible decision effects for candidate decision outcomes.  

Decision taking is positioned amidst many related notions including: decision making, decision process, decision making process, decision process making, decision engineering, decision progression, and decision progression production. 

\end{abstract}

\tableofcontents

\section{Introduction}\label{sec:Intro}
The objective of this paper is to conceptualize the notion  of a decision by making use of 
a number of techniques and concepts from informatics. In particular a decision will 
be defined as a tuple of components of various types, which is a common style of 
introducing a concept  in informatics.%
\footnote{For instance although the notion of an automaton sounds familiar to 
most people, in  informatics an automaton will be defined to constitute a tuple 
of interrelated items of various types.}
 This leads to a so-called constructing definition of ``decision" in the 
 terminology of \cite{Bergstra2011c}. I will propose to view decision making  
 as an informatics competence for the simple reason that  the decision 
 outcome primarily constitutes a piece of information.  
 I will use the terminology as well as the definitions of \cite{BDV2011b} concerning 
 competence and ability. That terminology includes the notions of a framework 
 competence, a community confirmed competence, an evidence based ability, 
 a  competence profile, and of a  conjectural ability. 
 Each of these notions will require instantiation to the decision taking context.
 
 Readers who have read and reflected upon Sections \ref{PAoD} and \ref{PSaSO} below, 
 will be said to have acquired a decision taking framework competence. 
 This puts them in the intended audience for taking notice of the constructing 
 definition of decision as given in Section \ref{Defdec}, as well as the subsequent theory development. 
 According to  \cite{BDV2011b}  a competence profile consists of a number of community 
 confirmed competences, which may be but need not be evidence based, 
 possibly augmented with one or more evidence based abilities lacking 
community confirmation, and possibly augmented with one or more conjectural abilities, 
which are lacking by definition of being conjectural both an evidence base  
and community confirmation. The conceptualization of ``decision'' below, 
together with some of its consequences qualifies as a ``theory of decision taking'', 
which is admittedly incomplete and of a limited scope. 
In addition I will provide specific proposals concerning conjectural abilities concerning decision taking 
which can be considered plausible consequences  of a person's awareness this theory.

In \cite{Nutt2011} the observation is made that decision making research 
mainly stands on two feet: description and prescription.%
\footnote{A similar dichotomy underlies the survey \cite{FallonButterfield2005}. 
That paper distinguishes descriptive and normative realms of (business) ethical decision making.}
 Description calls for observation and invites the development of theory 
 which may or may not be confirmed by observation. Prescription may lead to 
 improved behaviors not previously observed, or to a more frequent 
 occurrence of best practice. Prescription should be analyzed from 
 the perspective of a theoretical framework that supposedly explains 
 why prescribed patterns (of decision making) are to be preferred. 
 Nutt \cite{Nutt2011} indicates that action theory combines description and prescription.

I will pursue construction as an alternative approach to concept analysis which is 
standing on an equal footing with description and prescription. 
Decision is thought of as an idealized concept,  
found by means of a constructing definition, independent of any empirical observation, 
and without any intention for being prescriptive. 
The constructing definition of decision can be used by readers who are willing to forget 
about their prior intuition about decision, and to reconsider the notion in 
such a way that unexpected consequences may result, 
that is some events that were not thought of as decisions may 
turn out to be regarded as decisions and conversely.

\subsection{Disambiguating process product ambiguities}
This paper may be considered a case study in the disambiguation of a so-called process product ambiguity.
Many important terms and phrases somehow refer at the same time to the result of a process as well as to the very process leading to that result. According to~\cite{TallVinner1981} such ambiguities are visible already in
elementary mathematics. Remarkably, according to~\cite{GrayTall1994} resolving of such ambiguities,
which can be done by choosing one of several meanings as a preferred meaning for a term, is not necessarily a step forward in mastering a theme. 

Nevertheless I hold that:
(i) at least in some cases process product ambiguity stands in the 
way of a better understanding of a topic, 
(ii) the term (concept) decision is amenable to process product ambiguity,
(iii) ``decision'' is a term for which it is justified to look for a disambiguation, and 
(iv) disambiguation of ``decision'' in favor of an an action rather than the result of that action is most promising.

A similar disambiguation has been proposed for the notion of ``outsourcing'' in~\cite{BDV2011b} and~\cite{BDV2011c}. 
It should be stressed that there is no way that semantic ambiguities can be removed from one's language, 
not even from a formalized language tailor made to a specific topic. However, in some cases disambiguation of a 
notion that had become ambiguous may constitute a step forward. 

Evidently each proposal for disambiguation of a term (concept) will meet intellectual 
challenges related with its consistency, and besides those 
 challenges concerning acceptance. Indeed if no-one agrees
with the proposed resolution of an ambiguity the pragmatic value of the proposal is very limited.  
Various trade-offs will determine whether or not a specific proposal for the resolution of an instance 
of process product ambiguity is viable. 

For this paper, as well as for my project on decision taking based on this paper, it is only the academic 
challenge of developing a coherent framework (called OODT for outcome oriented decision taking) and of assessing its advantages and disadvantages 
in principle that matters.

Except for attempts to apply the framework of OODT in my own practices, no form of social engineering aiming at acceptance of the particular framework proposed below is among my objectives or ambitions.

\subsection{Getting started on decisions and decision making}
Writing a paper from first principles on decisions and decision making is a problematic objective given the 
amazing size and scope
of the existing literature on these themes. In \cite{Bergstra2010a} 
I have commented at length on this kind of difficulty, in the context of writing on money. 
Similar comments may be brought forward in the present context, 
and instead of repeating that part of \cite{Bergstra2010a} I will
only mention its existence. According to a conclusion drawn in \cite{Bergstra2010a}  
a best effort to write about a subject is legitimate even if guarantees of novelty cannot be provided due to the 
sheer size of the prior art. In addition guidelines on how to work properly in 
such circumstances are proposed.\footnote{%
Achieving compliance with these demanding 
guidelines is not an easy matter, however, an experience I made when 
writing on the concept of money in \cite{Bergstra2010a}, 
and again when writing on a specific non-classical logic in \cite{Bergstra2011a}. In \cite{Bergstra2010b} 
I made an attempt to explain why it is plausible to test control code, 
a position which I could not find anywhere in the massive literature 
on software testing which invariably takes the rationale of testing for granted. 
Confronted with the difficulty to grasp a massive literature on software testing, 
I made use of what I termed an informal logic. 
But I failed to notice that the notion of an informal logic is quite well-established 
already and that it has an entire journal devoted to it, 
something which clearly should have been made mention of in the same paper.  
References to papers on informal logic were subsequently provided in \cite{Bergstra2011a}. 
However, the guidelines from \cite{Bergstra2010a} just mentioned
imply that a new version of \cite{Bergstra2010b} should have been produced, taking these references 
(say \cite{Pinto2009,Waller2001,WaltonMacagno2010}) adequately into account.
Doing so, however, would violate the document uploading rules of the 
repository {\tt www.arXiv.org} which imply that only major modifications 
of a paper justify posting new versions. 
That leaves one with the question when inserting a previously missed 
reference is a substantial change of a paper, a question which seems not to have a uniform answer.}

In spite of the existence of a very extensive literature on decisions and decision making I 
will not start this paper with giving a survey on decisions and decision making for the 
simple reason that doing so adequately is a formidable challenge in itself and because it 
seems not to be a prerequisite for this paper. 

\subsection{Organization of the paper}
In Section \ref{PAoD} a number of issues are raised each related to decision. 
In  Section \ref{PSaSO}, preliminary design decisions are developed concerning a definition of decision. 
Section \ref{Defdec} contains my proposal for a constructing definition of decision. 
Subsequently and on the basis of that definition a number of derived 
concepts is developed and various ramifications of the definition are considered. 
In Section \ref{DTT} initial steps towards the development of a theory of decision taking are made. 
Decision taking theory includes decision quality, decision free management, 
implementation of decision mechanisms, decision making models, 
levels of decision making and taking, and decision taking in a structured hierarchy. 
Following the conceptual structure of competences and abilities of \cite{BDV2011b} a 
survey is given of conjectural abilities on decision taking which I 
believe to result from taking the constructing definition and other aspects of decision 
taking theory into account. Finally I return to the title of the paper by listing various 
informatics perspectives on decision taking.

\section{Preparatory analysis of decision}\label{PAoD}
Many aspects concerning the common intuition of a decision need to be contemplated before writing a constructing definition can take off. The constructing definition to be developed will take some aspects into account. In this Section I will consider the notion of a decision from different angles thus closing in on a preliminary survey of issues that may need to be covered by a definition of decision. 

\subsection{Primitive and non-primitive concepts}
Constructing definitions, by definition of that very notion, do not produce primitive concepts, that is concepts requiring no further constructing definition. But every constructing definition will ultimately rely on the use of primitive concepts if its author is committed to avoiding an infinite regress. These matters will first be considered in some detail.

\subsubsection{Decision: not a primitive concept}
Attempts of giving an extensive, detailed, or even informative definition of the concept of a decision are useless if decision is in fact a primitive concept that cannot be reduced in a useful way to other more primitive concepts. 

``For instance ``meaning'' might be considered a primitive concept, if one appreciates that defining the meaning of meaning without somehow making use of that very notion is difficult. Similarly the word important is not easily defined without already knowing what it means. I hold that the term choice is primitive in the same sense. 

If one assumes that a decision is a choice that matters for the future of the agent making the choice, then decision is what takes place if an agent (decider) makes a choice. If the agent is human, and if one believes in the existence of a free will, one may  insist that the agent's decision is a  manifestation (or an expression) of the agent's free will. 

I will assume a much larger distance between decision and choice, however, leaving choice a primitive concept which may be explained in terms of mathematical theories of an axiomatic nature, such as modal logic, process algebra (see \cite{BaetenBastenReniers2009}), whereas decision is a constructed notion that admits a reduction to (decomposition in terms of) a collection of primitive elements. In specific cases, the occurrence of choice may be among these elements.

\subsubsection{Decision: not an almost primitive concept}
Suppose that decision is identified with choice in the sense that each decision is a choice but not necessarily the other way around. In that case decision is not primitive because it can be reduced to another primitive concept. But the reduction is trivial, as it is a mere renaming which takes into account some additional requirements. Let a concept be considered almost primitive if its reduction to a primitive concept is a mere renaming which may express that some 
additional requirements have been taken into account. Thus ``occasion'' (with the intended meaning of a used car that is for sale) is an example of an almost primitive concept, assuming that car is a primitive concept. Now the following question arises: is decision an almost primitive concept. 
I propose that this question has a negative answer.

\subsubsection{Decision outcome: an almost primitive concept}
The existence of a decision requires more than the presence of the outcome of some process that must have involved a choice. Concerning the outcome a precise terminology will be needed.
In \cite{Rosati1981} (proposition 4) 
one finds that ``a considerable gap usually exists between the formulated decision and its implementation''.  This corresponds with the terminology proposed below as follows: decision outcome will be used instead of formulated decision, and effect is used instead of implementation. 
A decision outcome is a result, and the term result will not be specified any further.  As a consequence ``decision outcome'' is taken to constitute an almost primitive concept.

\subsection{Decision versus decisive action}
Some actions seem to require preparatory decisions and some decisions seem to be of decisive influence. Neither is by necessity the case, however. This leads to the following observations.

\subsubsection{Decisions cannot be postulated in hindsight}
In \cite{McConnell1979} one finds decision as an indication of an activity for which it is unlikely that it can be performed without a decision having been taken in advance in such a way that putting its outcome into effect involves or implies just that activity.  For instance a decision to state one's critical viewpoint in a public meeting may be assumed to have been taken in advance of an event (public expression) of that form. However, taking for granted that the existence of such a decision can
be postulated is unwarranted.

This use of the term decision as referring to a postulated decision  will be avoided below because it may 
lead to confusion and ambiguity. Can one talk about the early retirement decision, or the coffee making decision, or the emergency evacuation decision. In the latter case one might be tempted to state under certain conditions  that the  emergency evacuation decision has not been properly taken to express the fact that there was no such decision although there should have been one.

The term emergency evacuation decision (that is the postulated decision assumed by an external observer of an emergency evacuation to have taken place in advance and by which the evacuation was caused) is as problematic as the well-known top of a stack: what if the stack is empty. Then talking about its top postulates the existence of an object which fails to exist. 

\subsubsection{Decision taking need not be decisive action}
In \cite{Rosati1981} USA presidential decision is put forward as being part of a political process without decisive influence by itself. I will follow this view and assume that a decision need not be a decisive action, where a decisive action is characterized by being of major explanatory value in hindsight for the occurrence  or non-occurrence of a
family of subsequent events. 

One may wonder to what extent decisive action must be the consequence of decision taking. Or stated differently to what extent decision taking is a prerequisite for decisive action. This question has become very prominent in political science, because of the fact that several very serious human rights violations in the 20th century, each of which which can be considered to comprise decisive action, 
seem not to have been preceded by a well-organized decision making process.
Thus decisive action need not have been caused or preceded by a decision in spite of the fact that common use of
 language suggests the existence of a corresponding postulated decision. Indeed, the very notion of a postulated decision has been discarded already. Decisiveness cannot be assessed in real time (that is during or 
 immediately after an decision has occurred) as it is a labeling in hindsight only. In contrast decisionness (or degree of decisionality see \ref{Complications}) can be assessed in real time, at least in principle.

In other words decision taking need not be the taking of decisive action, but it may be. For an action to be decisive it needs to have significant impact. One may intend an action to be decisive, or expect it to be, but only actions from the past can be qualified with a significant degree of certainty, as having been decisive. 

\subsection{Deciding is decision taking}
Decision taking constitutes the culmination of a decision making process.  
The so-called decision process is more comprehensive  than the decision making process and it covers all of the  decision making process, and in addition to that the activities needed, or deemed necessary and for that reason applied, to facilitate or to bring about the decision making process. Decision making includes decision taking and in addition it comprises only but precisely that part of a decision process which directly influences decision outcomes. 

Decision process making, refers to the activity of one or more agents who are the driving forces behind a decision process. Decision process making can be either internal or external. In the latter case I will speak of external decision process management, which includes external decision process making consultancy. A decision process making agent manages or organizes a  decision process. A decision process may contain actions many of which are connected with the communication between agents who are directly or indirectly involved in the decision. A decision process may often be viewed as the putting into effect of a specific protocol, meant for a particular kind of decision. This putting into effect may be a multi-threaded concurrent activity 
driven by several agents, among which the decision takers, various decision makers, and some decision process 
makers as well as assistants for each of these.

The picture I will suggest is that systematic (perhaps professional) decision taking is embedded in equally systematic decision making. Decision making requires control both in terms of putting protocols into effect and in terms of protocol design. Meta-decision taking may be coined for taking decisions that occur within these tasks. Perhaps a better phrase is: decision process planning and running. This functionality is responsible for the concurrent effectuation of decision threads, as well as for adequate thread creation.

\subsubsection{Alternatives to taking a decision}
Deciding is always an action that exists in a context where agents have other options. Besides taking a different decision about the same issue, or taking an alternative decision about an alternative issue, 
at least the following options  for conducting an alternative action can be distinguished:
\begin{itemize}
\item [administration] Taking decisions that control the work of many other agents in a large variety of circumstances. Administration is  the part of an organization or institution most focused on deciding. Administration includes policy making about decision process design and control.
\item [networking] Connecting to other agents with shared interests.
\item [managing] Telling agents and groups of agents what to do, coordinating, planning, supporting, 
and monitoring their action.
\item [organizing] Designing patterns of activity, and managing agents to act in these patterns.
\item [political action] Influencing individuals and groups towards certain ends. Political action has the flexibility of being less procedural than decision making.%
\footnote{A political decision is a decision that takes place as a part of a political process, just as a business decision takes place in a business process.}
\item [operation] The lowest level of action where the work takes place. Operations rarely (and preferably not) interrupted by decision taking.
\end{itemize}

\subsubsection{Surveying agent roles in connection with a decision}\label{NDA}
Seen from the viewpoint of a decision a participant in decision taking may operate in different roles. Here are some options:
\begin{itemize}
\item [take] An agent may be taking a decision, individually or as a member of a group with  other agents.
\item [influence] An agent may (try to) influence the outcome of a decision.
\item [await] An agent may await a decision on which it fails to have any influence.
\item [trigger] An agent may cause a decision to be taken without having an influence on its outcome.
\item [request] An agent may ask for some decision to be taken.
\end{itemize}

\subsubsection{Temporal aspects of decision}
When speaking of decisions one may speak about future decisions about current decisions and about past one's. The fundamental flexibility of the concept of a decision is that it can be turned into abstract versions by leaving out information. With that in mind and allowing flexibility in timing, that is abstraction from time, 
the following statements and questions may be meaningful.
\begin{itemize}
\item [who] Who took (or will take) that (referring to substantial but incomplete information) decision.
\item [why] Why must that decision (that is a decision matching those specifications) be taken.
\item [when] When has that decision been taken, or when will that decision be taken.
\item [cause] Who (if anyone) took the decision to make that decision.
\end{itemize}

A decision ends a phase of indecision. A decision completes a phase where an audience of agents awaits the decision. A spontaneous and unexpected action by an agent is not considered a decision.

\subsubsection{Can decision be freely defined?}
When contemplating the various possible definitions of decision, I will entertain the hypothesis, if not the phantasy, that while decision can be assigned a meaning with some degree of freedom, many other related terms like action, event, process, thread, result, etc. have been provided with a known meaning which is not going to be changed in the course of a conceptual clarification effort on decisions. Indeed, dissatisfaction about an account of decision may in principle be resolved by changing the meaning of many surrounding terms, but going ahead in that way is not the idea of giving a constructing definition of a notion. 

The constructing definition of decision is developed on the basis of some preliminary understanding of decision, 
which is assumed to be available in advance. By taking notice of Sections \ref{PAoD} and \ref{PSaSO} of this paper a reader may acquire so-called framework competence (see \cite{BDV2011b} for that notion) concerning  the concept of decision which places him properly in the intended audience of the paper. In informatics the systematic definition of a concept labeled with seemingly familiar terms is often carried out under the heading of formalization, the use of that term being justified by the formal and mathematical appearance of texts. Besides formalization, however, the more important aspect of such definitions is to accompany an existing intuition of a concept with a much sharper, if not better, picture of that same concept. In giving the sharper definition a move may be made from a better picture of the (same) concept to an improved and modified view of the (adapted)  concept, which is meant to replace the previous view.

As a mode of working this may be considered quite arbitrary indeed, but I see no alternative path.

\subsection{Decision: like collision and unlike inscription}
An overwhelming majority of papers on decision making has been written by authors who apparently 
assume that the 
meaning of the phrase ``decision making''  can be taken for granted and for that reason  requires no explicit attention, 
at least not in their own paper.%
\footnote{An example is \cite{MellersSC1998} and also \cite{HeathfieldW1993}. Another example is \cite{Keeney1994}, and also the quite philosophical \cite{Dyckman2010}. The style of writing of \cite{Keeney1994} is quite common: decision making occurs in situations, which may be considered opportunities for the decider. Such opportunities may be recognized and identified, and rather than thinking in terms of the generation of alternatives decision taking or making agents must focus on the generation and survey of values which
determine what objectives are to be reached. In \cite{Keeney1994} DM stands out as a well-known and ubiquitous phenomenon admitting a flexible range of descriptions and analyses, without the risk that these make no sense 
because of a commitment to a narrow and specific definition of  DM.}
In this common understanding managers, consumers, politicians, and doctors, each make decisions, 
a task for which they may appreciate evidence based support, and moreover a task that admits detailed investigation.

I will assume that a decision is an act or event of deciding. Comparable cases are: an action is an act or event of acting, an explosion is an act or event of exploding, a collision is an act or event of colliding, a transmission is an act or event of transmitting, a computation 
is an act or event of computing, an execution is an act or event of executing. Participation is an act or event of participating.

In contrast to these examples, however, a permission is the outcome of an act or event of permitting,  an edition is the outcome of an act or event of editing, a construction is the outcome of an act or event of constructing, a translation is the outcome of an act or event of translating, a transcription 
is the outcome of an act or event of transcribing, an inscription is the outcome of an act or event of inscribing, a definition is the outcome of an act or event of defining, and an emission often is the outcome of an act or event of emitting. A (problem) solution is the outcome of an act or event of problem solving. Although a simplification is often the outcome of an act or event of simplifying, a  complication is usually not an act or event of complicating nor the outcome of such an act or event. Rather complication abbreviates the phrase ``complicating factor''.
With pollution the situation is slightly more complex. Usually pollution is the outcome of a plurality of acts or events of polluting, but in some cases it refers to the consequences of a single event.  

The terms prosecution, prevention, intuition, function, and  prohibition do not fit either of the above schemes. Rather these terms stand for general structures or roles causing or enacting their instances. This kind of meaning is implausible for the term decision.

I propose that decision represents an act or event of deciding, rather than the 
outcome of such an act or event.\footnote{%
If one holds that decision features so-called state/action ambiguity then I propose to resolve that ambiguity in favor of action.} Having made this ``decision'' concerning the meaning of ``decision'' in relation to ``deciding'', the phrase  ``decision outcome'' is can be used to refer to the outcome of a decision.%
\footnote{The phrase decision outcome can be found in \cite{Hage1980}. In \cite{Hickson1987} the decision outcome is referred to as the formal decision. I will not comply with that convention. In \cite{BGGT1987} a decision process, or equivalently, a decision-making process is said to end in the final making of a choice, or in the ultimate decision of choice.}
The decision outcome is an object, perhaps a virtual one, which can last in time, whereas the decision itself is 
bound to an agent, an instant of time and a place and immediately becomes a part of history, often having its decision outcome as its most enduring historic account.%
\footnote{Clearly an alternative is to have decision stand for what I proposed to be named a decision outcome. 
Then a phrase is needed for the act of taking a decision. No obvious candidate seems to be on offer, however. I will equate ``a decision has been taken''  with  ``a decision has taken place''. I will also equate 
decision taking with deciding.}

Viewing a decision as an act or event of deciding, rather than as the outcome of such an act or event, cannot be maintained in every context. For instance a design decision is the outcome of an act or event of deciding about a design. A personal decision will usually refer to the outcome of an an act of deciding by a single individual.  Decision taking, provided it is distinguished from decision making in some particular case, constitutes a final and somehow highlighted  stage of decision making, including the last act from which the decision outcome results.

\subsubsection{Preliminary deliberation not required}
In many descriptions of decision making (for instance \cite{Simon2004}) it seems to be taken for granted of a decision that it concludes a phase of deliberation during which several different options (candidate decision outcomes) are compared and that the decision cumulates in making a choice between these options, the decision outcome being identical to the chosen decision outcome option. I see no need for this assumption, and I consider the existence of a deliberation phase to be optional. Of course it will be often the case that a decision involves making a choice between several options, and that the decision process involves some form of deliberation admitting a comparison of those options, but it is consistent to assume that the only alternative that has been considered in the process leading up to a particular decision concerning some subject was not to produce any decision (concerning that same subject matter, that is with the same or similar objectives) at all at the time of deciding. 

Taking  the meta-decision 
$m_d$ to make a decision, say  $d$ concerning a subject  $s$, with $d$ not yet fully specified but with $d$'s outcome in outcome type $D$, or with an outcome constrained by requirements $R$, is itself part of the very decision process of $d$. In order to state this matter properly one must be capable of speaking about a future decision as an action that complies with some specifications and which will be refined from these specifications during the decision process that leads up to that decision.

\subsubsection{Decision ends a phase of indecision}
I propose that each decision is preceded by a phase, however short in time, of indecision. That is some awareness of the need or opportunity that a certain decision is about to be taken must be present at least within the deciding agent. Thus a decision ends an episode of indecision. The phase of indecision may be merely a postulated phenomenon, however, because its existence need not be provable in hindsight from documents or other records.

\subsection{Atomicity and scope}
The slogan that a decision is an act%
\footnote{I will use act as a shorthand for activity.}
or event of deciding cannot serve as a definition of a decision unless deciding has been defined. It only serves as a requirement on how to use words and phrases. A decision need not be atomic, it may split in several subsequent acts or events.%
\footnote{In \cite{Folsom1962} (p. 4) one finds: ``..It is often hard to pinpoint the exact stage at which a decision is reached. more often than not, the decision comes about naturally during discussions, when the concensus seems to be reached among those whose judgement and opinion the executive seeks.'' In \cite{Vickers1986}, however, a unique decision moment is said to exist. After that moment the decision cannot be taken again.}
More specifically, a decision is a progression (see \cite{BergstraPonse2009}) of acts or events that together qualify as representing an agent's activity of deciding.%
\footnote{This view is consistent with \cite{Simon1965} although
that paper seems to identify the decision process with the decision making process, and \cite{Simon1965} proposes that 
a decision is a progression of the entire decision process. In contrast I will assume only that
a decision is a progression of the decision taking process. A decision making progression is more comprehensive and it may involve steps not included in decision taking, such as the taking of a preliminary decision and the commenting and subsequent reworking and resubmission to the decision making pipeline of an improved preliminary decision outcome. A clear example of a decision making pipeline is presented in \cite{FriedH1994}.}

The decision taking process is a part of the decision making process, which may be imagined as the run of a decision making pipeline. A structured decision making process contains those steps of the workflow leading up to a decision which
are explanatory for the final decision outcome. 

Yet more comprehensive is the decision process. If a decision making process involves a meeting for taking a preliminary decision, the production of the inputs to that meeting will be part of the decision making process, while reserving the room and corresponding catering is part of the decision process but is not included in the decision making process. The decision process includes the actions and events of the decision making process together with all supporting and enabling activity, including catering, AV preparation, printing and copying, process control and monitoring, security, and transportation.

Decision process making (equivalently decision progression making) is the making of a progression of acts or 
events of a decision process, without participation in the decision making proper. Speaking in terms of progressions (or runs) I thus distinguish: decision taking progression (equals decision), decision making progression, and decision process progression.

Thus decision making ends with a decision (a decision taking progression) which produces a decision outcome. Decision taking is performed by the agent who is deciding. Decision making (equivalently: a decision making progression) includes decision taking at its tail but comprises more preparatory steps if any are present. One may feel the need to speak of a decision making process making to denote the task of the agent who sees to it that decision making takes place, that is that a decision making progression is being produced. Instead of decision making process making I will simply speak of decision process making. Indeed in order to ensure that a decision making progression takes place many additional and supporting acts may be required which will beu subsumed under decision making because of the absence of impact on the decision outcome.

Decision process making may be compared with theater making, which is the making of a progression of acts or events of theater play (playing). 
Theater making is often performed by someone not actually playing him- or herself. This suggests that an external (second) agent might be involved in order to bring a decision about for some agent, comparable the role of a theater maker. One may think in terms of decision planning and control, with planner and controller different from the deciding agent. If the agent is a single individual this thought experiment coincides with the issue raised in \cite{Valdman2010}. A plausible term for that role is a decision consultant. The decision consultant, if present at all, sees to it that a decision occurs, or equivalently that decision taking takes place. 
Decision consulting and deciding operate simultaneously and interactively. The decision consultant is not involved in decision taking, because the consultant sees to it that another agent will take a decision. But as the consultant may be quite influential its actions may be included in the process of decision making.

\subsection{An intrinsic circularity concerning decisions}
If one intends to define the notion of an inhabitant of a country, one finds that for understanding a country as a social structure, one needs the notion of an inhabitant already. The notions of country and inhabitant must be defined simultaneously. A similar issue pops up when contemplating a definition of decision, or more specifically a decision 
taken by an agent (or a collective of agents)  $A$. I will now argue that the concept of decision involves a
circularity which I don't see how to remove. In fact in the definition of decision in Section \ref{Defdec} I will ignore
this circularity, thus leaving open the question how to find an improved definition that takes it into account in a more serious way. The circularity comes from the fact that it is implausible to assign an agent the role of a decider without understanding what a decision is to begin with. Some agents simply cannot play that role. But in the definition in Section \ref{Defdec} I will not impose any constraints on the agent acting as a decider in an event to be considered a decision.

Indeed for $A$ to be deciding it must be possible that $A$ takes any decisions at all. The power to take decisions is constitutive for the concept of an agent for which it makes sense to assert that it takes any particular decision. A decision outcome $x$  of a decision $d$ taken by agent/unit $A$ can be a plan, in which case it must be assumed that $A$ has the power to see to it that the plan will be put into effect.%
\footnote{According to \cite{Parsons1963} power can be understood as a medium comparable to money capable of ensuring that outcomes are put into effect. Power then constitutes a background mechanism within which an agent may be capable of taking some decisions and incapable of taking other decisions. Defining the power of an agent in terms of the decisions it may both take and be sure to see their outcomes being put into effect seems to be consistent with the analysis of \cite{Parsons1963}. 

$A$'s power to make decisions can at each moment be defined as the collection of decision outcomes that (i) it may plausibly put into effect, and that (ii) are decision outcomes of decisions
that it may take when operating according to the rules that have been set for the various types of decisions.}
 In other words a decision $d$ can only be taken by an agent $A$ if it has the power to take decisions from a class of (potential) decisions, say $C_A$ containing $d$. After the outcome $o(d)$ has been put into effect, $A$'s power to take decisions has potentially been changed, that is $C_A$ may now differ (having become say $C_A^{\prime}$) as a side effect of the implementation of the decisions outcomes. An obvious example of that state of affairs is found with an agent $A$ who is deciding to buy some expensive real estate, and who, as a consequence of putting that decision into effect, is losing almost all of its cash and who is  from that moment onwards in  debt. $A's$ power to take decisions that will cost money (or rather, the putting into effect of the outcome of which will decrease the amount of money in $A$'s disposal) has been significantly decreased.

\section{Problem statement and solution outline}\label{PSaSO}
The problem to be analyzed and from some perspective solved in this paper is: what is a decision? In order to assess the usefulness of an answer to this question, that is a candidate definition of the concept of a decision, I will list some questions about the notion of a decision the answering of which would have to be supported by a candidate definition under scrutiny. In particular the answer given should be helpful to assess the following questions about decision processes in specific cases, in particular in a context of organizational decision processes.

\subsection{Some questions and answers on decisions}
The answers provided  for the following questions will support and hopefully explain the choices (design decisions)
made in the definition of decision that will be  given in detail in Section \ref{Defdec}.
\begin{itemize}
\item Given a specific theme or area, for relevance for some organization. 
What terminology is to be used about decision making concerning that theme? 

I will distinguish: decision process, decision, making, decision taking, and decision shaking. Given those notions: which agents are involved in the decision process,  in decision making, in decision taking, and in decision shaking. 

Decision shaking is a political process performed by agents or groups of agents outside the hierarchical control of the decision taking agent, after and as a consequence of the decision having been taken. It leads to its destruction in hindsight as an authoritative statement. It renders the decision outcome futile. It may also have negative impact on the position of those who took the decision and even on those who were involved in the decision making 
process leading to the (shaken) decision.

\item Suppose that an agent has concluded that he will probably be involved in decision making, or in decision taking concerning a topic: which activities are  involved in that role?

An answer to this question depends on circumstances which may vary from organization to organization and from theme to theme. Given an organization and a coherent (that is interrelated) bundle of themes of comparable importance for the organization, it is reasonable to assume that decisions can be classified into a number of categories, such that for each  class taking a decision within that class requires that some protocol of preparatory actions (which may include some decisions) must be followed. Such a protocol is informative about the interface that an agent may provide and make use of.

\item How to refer to those activities involved in a decision process that are not part of decision making? Where are the boundaries with decision making and how to assess the relevance of these actions for the decision process?

To answer these questions an interface of basic actions (see \cite{BergstraPonse2009}) must be determined which indicates the activities that can be performed by the chief decision taker. Some of these actions comprise the issuing of impositions to other agents who act in supporting roles.

\item How to name key roles in the decision process? Here I will assume that human agents are at stake. If a group is taking a decision each of the members is said to be taking a decision.

I will speak of a decision taking officer (DTO), given an organization and a bundle of themes of relevance for that organization,  if an agent is regularly involved in decision taking about one or more of the themes in the bundle. A chief decision taking officer (CDTO) is always involved in the decision taking processes of highest importance. Both DTO en CDTO are supposed to have decision taking as their main activity within the organization. 

If a person is taking decisions only occasionally he is classified as DTP  for decision taking personnel. Every (C)DTO is 
also DTP. Decisions taken by DTP in that capacity are either final decisions, that is decisions constituting the termination of a decision process, or preparatory decisions that are part of the protocol leading to a  final decision. 

Besides DTP there is DMP, decision making personnel. Non-DTP DMP is at least occasionally involved in preparatory steps for decision making. DMP personnel not classified as DTP need not be taking orders only. They may act at their own initiative to contribute to various phases of decision making, for instance by analyzing the expected effect of a proposed decision outcome, or by analyzing the risks posed by unintended side-effects of a decision (that is of putting
into effect the decision outcome if the decision were taken), or by unintended consequences of a proposed decision outcome (equally hypothetical of course). Yet more comprehensive is DPP, decision process personnel. Non-DMP DPP has a supportive role only and will not take any influence on decision outcomes.

\item How can a dedicated decision taking agent be characterized in mechanical terms? In other words: what kind of procedures are put into practice by an agent with a primary focus on decision taking (at some level of abstraction and in the context of some organization)? 

The perspective of a (C)DTO is as follows: different decision processes take place as threads in a multi-threaded system. The (C)DTO is responsible for scheduling the multi-thread of decision processes by means of an appropriate form of strategic interleaving (\cite{BergstraMiddelburg2007}). In doing so the (C)DTO instructs other agents to take part in the decision process at large and in decision making fragments of it.

\item What drives a decision process?

A  decision process is a thread consisting of the putting into effect (see \cite{Bergstra2011c}) of a single pass
 instruction sequence P (see the program algebra outlined in~\cite{BergstraLoots2002}). A (C)DTO is putting initial segments of P into effect while regularly extending it as an outcome of a planning process. The mechanics of the planning process and subsequent
 thread extension are left unspecified in the formalization of thread algebra (see~\cite{BergstraMiddelburg2006,BergstraMiddelburg2007}).

\end{itemize}

\subsection{Meta-decisions on the definition of decision}
The proposed answer on the question what constitutes a decision as embodied in the definition in 
Section \ref{Defdec} below  involves some meta decisions (alternatively called design decisions) 
with which one may disagree. Without taking the risk of such disagreement writing this paper is pointless. Here are the main meta decisions that enter my explanation of what is a decision:

\begin{itemize}
\item A decision is an action occurring in space and time, 
and it is performed by an agent who is responsible for the decision. 
\item A decision has an outcome, which is a representation of the content of the decision in a form which can endure in time. For instance a digital text posted on a website controlled by the responsible agent.
\item The outcome must be distinguished from the effect of a decision, which is best seen as the consequences of putting the outcome into efect.%
\footnote{In \cite{WestbrookNT1978} one finds the convention that decision outcome stands for what  I propose to call the effect of the decision outcome. If it has been decided to buy some gadget, the actual purchase is caused by that decision, and it may be considered the effect of the decision outcome.}
There may not be any effect if no agent bothers to put into effect the decision  outcome. Producing the effect of a decision outcome by putting it into effect it is not part of the decision process of that decision. 
The decision process leads up to the generation and making available of the decision outcome and ends thereafter.
\item Decision taking differs from making a choice or expressing a preference, and decision making also differs from the determination of a preference.%
\footnote{Nutt \cite{Nutt2011} considers choice a possible unit of analysis occurring as one of many actions to be considered when having a decision focus. He distinguishes between the unit of analysis and the level of analysis. Nutt also mentions the nesting of decisions as an aspect of scope regarding decision making research. In \cite{Etzioni1988} decision and choice are identified, though not explicitly. In \cite{Saaty2008} and \cite{MahalelZK1985} decision and choice are treated without distinction. In \cite{CyertST1956} a decision is essentially a choice but compiling the menu of options as well as developing predictions of the effects of various decision outcomes (that is options under the assumption that these have been chosen) is considered part of the decision as well. In \cite{WallachFA2010} ethical decision making is understood as action selection under ethical constraints.} 
Such tasks are often input to a decision but do not constitute the decision itself. As a consequence of this meta decision a major part of the literature on decision making has to be reclassified as being about making a choice or optimizing a possible selection from a menu of options. 
Thus: choosing is not deciding about a choice. Choosing is more primitive than deciding and less context sensitive. Deciding, however, may be based on a choice. If a choice lies on the path (pipeline, workflow) which creates a candidate decision outcome in preparation of a decision $d$  to be taken, that choice constitutes part of the decision making process (and hence of the decision process) for the decision $d$ but not of the decision taking process.
\item Decisions are always taken in a context where some awareness of expected effects (of the decision outcome, 
that is of putting the decision outcome into effect) is present. Decisions are taken in order to bring expected effects of their outcome about.
\item A decision $d$ may itself be caused by one or more preparatory decisions which play a predefined role in a decision process. Taking the preparatory decisions is part of the decision making that culminates in the decision $d$ being taken.
Activities needed for the decisions involved in the decision making for $d$ are part of its decision process.
\item Even if a decision $d$ produces an outcome with the subsequent and intended effect $e$, that effect may not 
have decision $d$ as its most prominent cause. For instance it may be the case that some preparatory decision $d_p$
has given rise to a state from where unavoidably $d$ was going to be taken, in which case the preparatory decision$d_p$ is an original cause of the effect $e$ (of putting the outcome of $d$ into effect) rather than the decision $d$ itself. 

Nevertheless the agent $a$ who took decision $d$ may subsequently be held responsible for the effect $e$ (of implementing the decision outcome of $d$) even in the case that the ``real cause'' $d_p$ has been a decision taken by another agent, say $b$. Perhaps $b$ is held morally responsible for $e$ in such a case.

\item The term decidability as used in the theory of computation refers to the possibility to make some choice effectively, that is uniformly computed by means of an idealized computer.   In the absence of any intended effects of that choice, an instantiation of decidability as an act of effectively making a choice between several options must not be considered a decision, but merely a choice. A consequence of this meta decision about the meaning of ``undecidability'' is that the so-called undecidability of the halting problem (see \cite{BergstraMiddelburg2012}) 
is not about the absence of the possibility to make some decision but about the absence of a method for effectively making a choice. So I would prefer the phrasing that the halting problem is not effectively solvable.
\end{itemize}

\subsection{Alternative approaches for defining decision}
An obvious difficulty with the above requirements stated about the notion of a decision is that whatever definition one comes up with,  it will not fit in a few lines,  thus defeating the extreme conciseness which characterizes
concept definitions favored in the circles of management science. 

Trying to provide a shorter 
definition of a decision, which conveys some but perhaps not all of the content considered essential for a decision 
brought about these ``shorthand definitions'':
\begin{enumerate}
\item A decision is the promotion by an agent of some data representing a proposed decision outcome to the elevated status of a decision outcome.

\item An example of a decision is the act of giving (or refusing) permission to another agent to perform some activity, given a request by the other agent for that permission. The decision outcome is the statement (including motivation) of that permission (or refusal) in a durable form. 

Another example of a decision is to turn a preliminary decision into a final one after accommodating comments by various parties on the decision outcome of the preliminary decision. The decision outcome may for instance be the written intention  to release funds for certain purposes, or the written intention to organize a specific meeting, or the written intention to terminate some operation, and so on.

In principle the concept of a decision can be obtained as an inductively found generalization from a limited number of significant examples.
\item Decision taking (making) is the application (that is a meaningful instantiation) of a competence which one may prove to be in command of by having participated in a substantial range of decision taking (making) activities. Decision taking (making) competence is a community competence in the sense of \cite{BDV2011b}. 

Two forms of openness can be distinguished: open expectation, open intention. I am inclined to rank open decisions of either or both kinds higher than closed ones. In addition an open intention decision can be deceptive which diminishes its quality.%
\footnote{Of course there are circumstances where the deception coming along with a decision is its key quality, but I will consider those circumstances exceptional and I will not insist that the mentioned quality criteria will apply as well in such exceptional cases.}
\end{enumerate}
 
\section{Definition of decision}

\subsection{Decision defined by way of construction and  filtering}\label{Defdec}
The definition below comprises two parts. First a constructive definition introduces entities that might count as definitions. Having defined the kind of event which constitutes
a decision, the next step is to list constraints phrased in a negative form, filtering out events that on closer inspection are not considered decisions.\footnote{%
This definition of a decision outcome may be considered an extension (enrichment) of the definition of a promise.It follows that the core of a decision is a promise as defined in \cite{BergstraBurgess2008}, where a promise is taken to be a documented intention with an explicit scope of announcement.}

\subsubsection{Constructive part of the definition}
A decision is an event (below referred to as The Event) 
which is specified in detail as a tuple containing the following items. \begin{enumerate}
\item [time and place] Coordinates in space and time for The Event (with spatial coordinates being less important if the agent is an aggregate operating in a distributed fashion). The temporal information may provide a time interval rather than a single moment in time, because The Event need not be atomic in time. 
\item [timing] 
The timing mechanism indicates how, or under which constraints, time (and place) of The Event are found. 
Concerning the timing mechanism several scenarios may exist such as for instance:
\begin{description}
\item [real time.] The Event must take place at some specific moment in time. Its type and requirements are known in advance. Clearly human decision taking has an inherent imprecision in timing which cannot be overcome. Faster decision taking must be automated, which implies that some machine or software agent is considered to play 
the role of a  decision maker.
\item [fixed deadline.] A decision (of some type, and satisfying some requirements of its outcome) must be taken before some known deadline.
\item [opportunity interval.] The decision may need to be taken within a given time interval, that is a temporal window of opportunity.
\item [open end.] There is no firm deadline for a decision but its type and more  specific requirements are known.
\item [go, no go.] Like an open deadline decision but now the outcome requirements are fully specified. It may be that after some moment in time the outcome is vacuous, that is, the implementation of the decision outcome is an empty process if some moment in time has been passed.
\end{description}

\item [decider] The decider is an agent, including the option of an aggregate agent (that is a group of agents), in the role of a decision taker.
\item [agent role] A role in which the decider  operates with regard to The Event. This attribute is essential if the agent is acting in different roles simultaneously at the specified time and place. An agent role can only be determined with some reference model of a social structure or an organizational framework in mind. This attribute must provide a true role, and not merely a name for it. From a role the power of an agent can be derived, at least in principle.

\item [input data] A package of data playing the role of decision inputs. These inputs typically include, a classification of the decision outcome to be produced, together with protocol information on how that kind of decision is to be  made and taken, a menu of options, preferences imposed on the various menu items, test reports, artifact reviews and assessments, advice from external consultants, motivations for preferences, results of optimal choice analysis between menu items, proposed decision outcomes, historic data about the coming into existence of these proposals including data about who has been consulted on the basis of preparatory decision outcomes, which objections have been taken into account etc.

\item [outcome] A text, or more generally a meaningful symbolic or graphical code, in rigid  or in spoken form, stored with some permanency, playing the role of the so-called decision outcome.

A decision outcome is a description of a state of affairs which either is put into effect by the very decision taken,
or can be put into effect by subsequent action performed by a standing organization of agents or by 
self-organizing agents. These agents will invoke their own activity on the basis of their acknowledgement of the role of the decision taker.

\item [outcome type] A type or class to which the decision outcome is supposed to belong. The type information may also indicate which protocol must have been followed in preparation of the event.
This is a non-exhaustive set of possible decision outcomes types:
\begin{description}
\item [reply.] A Boolean reply concerning a question that has been put in advance.
\item [assertion.] An assertion of a fact, 
or of the endorsement of a fact. The fact may either be spelled out in the decision outcome or it may be known via a reference. Assertions may have further types such as: verdict, opinion, hypothesis, guess, claim, confirmation, and rejection.
\item [plan.] A plan to be put into action once triggered externally is a certain way.
\end{description}
\item [outcome novelty] 
Given a decision outcome type additional constraints (or rather constraint types) on the outcome may exist, constraining the novelty of the outcome. Such constraints constitute part of the decision. Here are some possible values for this attribute.
\begin{description}
\item [closed solution.] The decision outcome may be almost entirely known for some time already, and one waits for the corresponding decision to take place.  (In this case the outcome represents no novelty at all.)
\item [indication.] A question together with a menu of alternative answers may be provided. The requirement is that the decision outcome indicates a choice of an answer. For each choice some numerical parameters may need to be instantiated in addition.  (In this case the outcome represents no novelty at all, but the choice that has been made 
may be unexpected.)
\item [half open solution.] The menu of alternative answers as just mentioned may be merely an indication and other potential solutions may well exist. Then the decision will produce an outcome that may comprise the result of some creative activity. (This case implies limited novelty only.)
\item [open solution.] A problem area may be given together with the assertion that the decision (or rather the consequences of implementing its outcome) must contribute to its solution, though no indication of possible outcomes satisfying that requirement are given. (Novelty is possible.)
\end{description}
\item [protocol] The protocol indicates how the decision process leading to The Event must be shaped. At least the protocol
indicates the start of the phase of indecision which has been brought to completion by The Event.
\item [scopes] There are several scopes involved in a decision:
\begin{description}
\item{\em endorsement scope.} The endorsement scope contains those agents on behalf of whom The Event is performed. It contains at least the decider (who has been listed in a previous item). (Members of the endorsement group are alternatively called co-deciders or co-decision takers.)
\item{\em primary announcement scope.} Contains those agents to whom the decision outcome is addressed.
\item{\em implementation scope.} Contains those agents whose behavior will be guided (by those agents constituting the announcement scope) so as to put the outcome into effect.
\item{\em secondary announcement scope.} Those agents outside the primary announcement scope 
who will be told (or may be told) about the decision. This set may be empty in the case of a secret decision (also
called a hidden decision). It is assumed that the decider sees to it  that the decision outcome is not communicated outside the secondary announcement scope. 
\item{\em effect scope.} Those agents whose existence is supposed to be influenced by  the decision outcome
being put into effect.
\end{description}
\item [public expectation] An expectation of the effects that announcement of the (decision) outcome will have. This information is made available within the announcement scope. 
\item [public intention] An intention of the agent together with a motivation why the decision conforms to that intention. In particular it must be guaranteed that the decision taking agent has some grounds on which to base the expectation that the effect of announcing the decision outcome complies with the intentions. In an open intention decision the intention is documented and is communicated as a part of the decision outcome. In that case the decision may be considered an enrichment of a promise. In a closed intention (or secretive intention) decision, the intention is not communicated in the way mentioned above.

An open intention decision may be deceptive if, viewed as a promise it is a deception 
(see \cite{BergstraBurgess2008} for that notion in the context of promises).
\item [private expectation] An expectation of the effects that announcement of the (decision) 
outcome will have. This information is made available only within the endorsement scope. 
\item [private intention] A private intention is optional. If it exists it differs from the public intention.
\item [risk analysis] (Optional) a risk assessment that the decision outcome will fail to 
lead to the intended consequences. Risk assessment may split in a private and a public component.
\end{enumerate}

\subsubsection{Filtering out instances of choice}
The above constructive view of events that may count as definitions is in fact too wide. Negatively formulated constraints
are used to filter out events which might comply with the constructive definition but which I propose, nevertheless, 
not to count as decision.
\begin{description}
\item [Not a choice.] A decision outcome must not be
a selection made from a predetermined menu of options. An action leading to such a selection is a 
choice (act of choosing) rather than a decision.\footnote{%
A choice may constitute an essential part of decision making, because enacting a choice  may be needed in
advance of the composition of a candidate decision outcome. It a committee searches for a person to appoint for a job it will choose who to appoint from a ``menu'' of available candidates, and then a higher body will be advised thus, so that
it can take an appropriate decision, an action which does not involves making a choice anymore.}
\item [Not a voting.] Voting is a special mechanism for making a choice and for that reason it will inherit not counting as a decision from its superclass choice.
\item [Not an instance of action determination.] Action determination is the act of determining how to act 
immediately followed by performing that action.\footnote{%
Car drivers while driving perform action determination rather than decision taking.}
\end{description}
\subsection{Derived notions about decisions}
Having a precise definition of decision available a variety of notions can be developed on top of it. These notions are useful when speaking of decision processes, decision making progressions, and decision taking progressions.
\begin{description}
\item{\em Implicit decision.} An implicit decision (also named a postulated decision) is not  a decision, at least not in general. In other words the decision to $\phi$ need not have existed even if $\phi$ is taking place and some observers may think of a decision to $\phi$ as a necessary precondition for doing $\phi$. 

However, if an organization functions in such a way that certain activities, say $\phi$ can only be performed when based on a preceding decision to that end, that decision may be referred to as the decision to $\phi$, and such a decision may be termed an implicit decision
\item{\em Primitive decision.} A decision $d$ is primitive if it is not a consequence of putting into effect a previous decision $e$ for which the occurrence of  $d$ was an intended effect (that is such that the occurrence of $d$ features amongst the intentions that constitute $d$).
\item{\em Decision error.} Various errors can occur when  a decision is taken. For instance the co-deciders may not be 
sufficiently involved, an agent in the primary announcement scope may be missed out, agents outside the primary announcement scope may be informed. Further, the outcome may be phrased in meaningless language, it may be inconsistent with the public intention, time and place may not match the timing constraints, the agent may not play its stated role, the expectation (about the consequences of putting the outcome into effect) may be  unjustified.
\item{\em Decision management.}
Decision taking is performed by decision takers and co-decision takers, but it constitutes an activity which is mediated by other agents and tasks. Decision management is performed by agents whose task it is to see to it that useful decisions are made properly. It is possible but not necessary that decision takers are decision managers as well. 
\item{\em Decision orchestration.}  Decision orchestration constitutes an aspect of decision management with a focus on the design of individual decision processes. Decision orchestration involves questions like: who must take a certain decision? What protocol must be involved when taking a certain kind of decision? Who should be involved in decision making, given a decision taking protocol. 
\item{\em Decision choreography.} 
Decision choreography may also be considered a branch of decision management.
Decision choreography takes place if a variety of decision processes is to be managed in parallel. Its focus in on the arrangement of interconnections between different decision processes that are progressing in parallel.
\item{\em Decision taking competence.} A community competence emerging from having played a variety of roles in a variety of decision taking processes (see \cite{BDV2011b} for a description of community competence). 
\item{\em Decision making competence.} A community competence emerging from having played a variety of roles in a variety of decision making processes. 
\item{\em Decision ratification.} In some cases a decision once taken needs some kind of public confirmation, for instance: crowning a king, handing over a certificate, publicly announcing an agent's bankruptcy. Ratification may be used for this, though in some cases ratification still may fail so that it is closer to decision taking after all.
\end{description}

The use of the terms orchestration and choreography has been borrowed from service science where these terms have been used with considerable success with a clear technical meaning from which the above proposals have been derived (see \cite{Peltz2003}).

\subsection{Ramifications arising from the definition}\label{Complications}
The definition of decision given above gives rise to further questions. These questions may indicate the need for modifications and refinements of the definition.
\begin{description}
\item{\em Decisionness.} In spite of the lengthy definition just given, it seems to be the case that an utterance of an agent being a decision is a gradual matter. There are no definite demarcation lines. Rather than defining when an event constitutes a decision, one may understand the above definition as an outline of a description of the concept of ``degree of decisionality'' (that is the degree to which the event qualifies as a decision).%
\footnote{In \cite{Bergstra2011c} I have introduced the notion of a degree of executionality in order to deal with the problem that I could not find any convincing and straightforward definition of the notion of instruction sequence execution (a progression of machine steps constituting an execution of an instruction sequence, with instruction sequences defined as in 
\cite{BergstraLoots2002}). In \cite{BDV2011c} the degree of outsoucingness was coined in order to deal with gradual phenomena that occur if one plans to define under which circumstances a sourcement transformation qualifies as an outsourcing.}
If all attributes are present the degree is very high (say equal to 1). If only an outcome exists and all other attributes are absent the degrees takes its lowest value, say 0. Different groups of stakeholders may disagree in their decisionality assessment of the same event. The mere statement that an utterance is 
referred to as  a decision by the agent making the utterance does not in itself contribute to the degree of decisionality, though many agents will be happy to label their own utterances as decisions, in spite of defects concerning one or more of the criteria mentioned in the definition of decision above.
\item{\em Definition complexity.} The above definition of  decision is both lengthy and complex. Can it be the case that an important notion like decision is in need of a definition of this complexity? Or is the entire project of defining decision 
heading in the wrong direction if it leads to a result that is hard to memorize in the first place. For the moment I think that a definition of this complexity may be needed to find firm ground for a theory of decision taking. More concise definitions can be developed subsequently for application in a specific context.
\item{\em Decision templates.}  When it is said that some decision must be taken, this means that partial information about a decision is given in advance (called a decision template), which may or may not include the identity of an agent,  that it is expected of some agent (or in case it is contained in the decision template, the agent mentioned in the template) that it produces a decision the the description of which completes the given decision template. Many different decision templates are conceivable, and much less information than required by the above definition may be included in a template. The concept of a decision seems to have been simplified by the simplicity of its most common templates. Notwithstanding that, a decision when taken brings together all information as mentioned in the above definition.
\item{\em Nested scopes.} A decision can be unexpected for agents in its scope. To understand this it must be assumed that scopes are collections of agents, and that various scopes, ordered by inclusion come into play when defining a decision. 

For instance, the phenomenon of an episode of indecision will be noticed by agents in some scope between the 
endorsement scope (decision taker scope, or decider scope) and the primary announcement scope. The scope of decision making agents (decision maker scope) extends the endorsement scope, but it need not be included in the primary announcement scope. 

\item{\em Label justification.} Is every event which is called (labeled) a decision in fact a decision. This is a matter of justification. And conversely: are all decisions (that is activities or progressions that comply with the above definition of decision) indeed marked as decisions. The state of affairs seems to be as follows. Labeling a progression a decision may be unjustified. 
But agent $A$ may still have an interest in doing so. A decision need not be labeled (called, referred to) as a decision. But some actions may be elevated to the status of a decision by being referred to in that way.

If in a religious ceremony the minister or priest declares a couple married, that act qualifies as a decision according to the given definition, although it will not often be labeled as such. The couple has made a preliminary decision, the effect of which has been amongst other effects that the dignitary has planned the ceremony and has prepared the certificate that serves as a decision outcome. What makes one reluctant to label the minister's or priest's action as a decision is the lack of choice. The dignitary, however, might have declined to marry the couple for a variety of reasons. 

\item{\em Causal chains of decision.} Given an organization, some activities performed by individuals or groups of its members require that some preparatory decision has been taken so that the activity can be understood as a consequence of the decision. 
\item{\em Non-decision actions.} Most actions that occur in a business process are not part of a decision process. This matter has been discussed in Paragraph \ref{NDA}. Once a focus on decisions is introduced and a decision is said to be an action (which may or may not be atomic in time and space), the term action is not by default referring tot a non-decision. Lacking a positive qualification for an action not being a decision I borrow the phrase target action from \cite{BergstraMiddelburg2007}
for non-decision process actions. The idea is that decision making is performed within a system in order to allow it to
properly perform its main task consisting of a multitude of target actions. For instance the decision to schedule a course leads to the many target actions involved in delivering the course.

The notion of a target action is relative to a class of decisions. Seen from a higher level of abstraction, that is from the standpoint of decision choreography the decisions occurring in a progression of one of the participating decision processes are mere target actions. However, at the level of a particular decision process resulting from decision orchestration, the same progression may be viewed as an alternation of decisions, decision making actions, decision process actions, and target actions. These target actions may in turn be decomposed in an choreography  controlled parallel composition of orchestrations each of which may constitute an alternation of target actions (at some lower level), decision making actions and decisions. 
\item{\em Can animals take decisions?} The definition given above leads to the proposition that animals cannot take decisions, mainly because they cannot produce ``results''. Here a result consists of storable and meaningful information. Of course most animals can make choices, but that is a different matter. Still this assertion is a matter worth more attention, it might be mistaken on biological grounds, or it might be considered unfortunate to the extent that it constitutes an incentive to rework the definition  of decision.

Once this view is adopted it becomes implausible that a solitary operating human being without support of some form of technology, including social and organizational technology, can take decisions.
\end{description}

\subsection{Demarcation from neighboring concepts}
Ins spite of the detailed definition of a decision given above the concept of a decision admits a continuous transformation into several neighboring concepts. Here are some gradual traditions to other notions from a decision which may occur:
\begin{description}
\item [Decision outcome.] Under gradually changing circumstances the distinction between decision and decision outcome may evaporate, with as a consequence that the process product ambiguity of decision reappears. 

This gradual transition may for instance  occur if the action involved in a decision becomes less and less pronounced, perhaps undetermined in space and time, an a situation may arise where the decision outcome still exists but no act of deciding counts as the origin or cause of existence of the decision outcome.

\item [Promise issuing.] The issuing of a promise may become indiscernible from a decision in each of the following (continuously changing) conditions: if the role of the decider becomes less pronounced, if the decider is increasingly inclined to play a role in the effectuation of the decision outcome, if the decider is increasingly disinterested other agents packing part in putting the decision outcome into effect, if the decision outcome implies some form of obligation on the decider, etc.

\item [Choosing ands voting.] In a family of related conditions for decision taking the decision outcome may increasingly be considered a choice from a menu. This vector of change of a decision allows a continuous transformation into a choice or, if the appropriate mechanics for choosing are in place, a voting.

\item [Adopting a plan.] A plan may be consider a promise an agent issues to itself. 
Adopting a plan may be considered a form of issuing a promise to oneself. in just the same way as the demarcation between deciding and promise issuing is fuzzy (that is, there is no sharp demarcation), the demarcation between deciding and plan adoption is fuzzy.
\end{description}

\section{Terminology for choice and planning}
Choice is unlike decision in that no choice outcome is expected preceding the effectuation of the chosen activity.
Choice involves making a choice between given alternatives for further action. Choice unlike decision involves
a low degree of design.

\subsection{Choice} I will propose a terminology for choice that is somewhat remote from conventional usage
in order to have a better tuning with the proposed terminology for decision.
A choice is an act of choosing. For a choice to occur (as an action performed by some agent $A$) at some stage
$A$ needs to generate a menu of options for further action from which the choice is to be made. 
Generating the menu is a part of the choice preparation phase of choice making.

Once the menu is available to the agent $A$ a sequences of steps may be 
required for $A$ to find out which of the actions from the menu will be chosen. 
A choice issuing thread, (or choice taking thread) may have been
defined during the choice preparation phase and effectuation of the choice taking thread by $A$ will end in
some of the actions from the mentioned menu being performed. 
The act of performing that action indicates that a choice has been issued.

A formal choice outcome may be considered as being a description of the choice (the chosen menu item) 
that has been issued. Choosing, however does not entail the production of a choice outcome, 
it merely entails the effectuation of the choice outcome.

If instead of effectuating the choice outcome, a mere description of the chosen menu item is produced by $A$ as an outcome a plan has been produced rather than that a choice has been issued.

Choice determination refers to factors which may allow an external agent to predict the selection that agent $A$ makes when issuing a choice.

\subsection{Plan issuing}
Planning is plan making. Once a plan has been prepared and plan making has come to an end the final stage is 
the production of a plan outcome (usually simply called a plan) and the inclusion of the plan into an agent's portfolio of plans.

I hold that after the plan has been issued an agent still needs to issue a choice between 
different plans as a precondition for the effectuation of a plan to begin. 

It is plausible that choice making involves an extensive planning phase in which several plans are issued
together constituting a menu from which at some stage a choice is made. 
Issuing the choice may entail an announcement to other agents of the plan that has been 
selected as a first step of its effectuation.

\subsection{Action determination}
I will use the term action determination for a process that in real time brings an agent to finding what to do,
in terms of activity, and doing it. The difference between choice and action determination is that in the latter no choice from a prefabricated menu is involved. Admittedly there is a somewhat unfortunate terminological collision 
between action determination (done by the acting agent) and choice determination (done by an external observer 
of the acting agent).

\section{Decision taking theory}\label{DTT}
The extensive definition of decision may prove its value by constituting a productive point of departure for developing a theory of decision. The definition itself must be considered a part of a theory of decision, to some extent it already qualifies as a theory of decision, irrespective of the merits of that theory. 

Decision theory is a classical phrase. In \cite{Suppes1961} and in \cite{Jeffrey1956} it is identified with the theory of making a choice between a variety of possible actions. It is possible to leave that meaning unchallenged if one admits or accepts that decision theory is not about decision taking but about choice making (that is choosing). For this reason I will speak of decision taking theory if the theory is about decision taking with decision defined as in Section \ref{Defdec}. Of course decision taking theory has many variations parametrized by different definition of decision and decision taking. Nevertheless I require of decision taking theory that it is based on a concept of decision which is takes a choice as an input rather than incorporating the choice as its essence.

\subsection{Leadership without decision mechanism: decision free management}
The language of decision making has become so ubiquitous that it seems obvious that organizations must have leadership for taking decisions. This is clearly not true, however, because an organization can for instance be managed by agents who issue commands which are not resulting from any known or specified, let alone monitored, decision process. It is also possible to manage an organization by having informal rules in place which to some extent allow subordinates to find out what they should do or say in order to please their leadership. In that scenario the leadership may confine itself to issuing rather vague and abstract declarations only, nowadays often called mission statements, which are subsequently interpreted by functionaries working at a lower level in the organization.

Decision free management is an effective form of management for relatively small organizations.%
\footnote{In The Netherlands, from some size onwards, an organization needs to maintain a works council operating according to a Dutch law, the WOR (wet op de ondernemingsraden). In an organization that uses decision free management in the absence of a works council, the introduction of a works council may necessitate putting well-defined decision making protocols in place at various levels of the organization. A works council interacts with management in terms of a discourse about decisions, decision making and decision outcomes.}

Decision free management may operate in many different flavors and styles, and may adapt itself statically or even dynamically to different circumstances. The two extreme forms are a top-down line of command and a loosely coordinated collective of functional agents striving towards a common abstract goal. Each decision free management style can be modified, and sometimes improved, by introducing some forms of decision taking.

\subsection{Implementing a decision mechanism}
For a decision to occur or to be taken by some specific agent or by some group of agents certain preconditions have to be met. Getting these preconditions arranged amounts to implementing the very concept of a decision itself. 

Thus implementing decisions takes place at a different level of abstraction from the implementation (putting into effect) of specific decision outcomes. In order to highlight that difference I propose to speak of ``implementing a decision mechanism'' rather than of the equivalent ``implementing decisions''. Boards, management teams, councils, congresses, and so on each provide such arrangements in different ways. I understand the ubiquitous presence of management teams and boards of directors as an indication that a decision making structure has been put in place. Without any such structure decisions cannot be taken, though perhaps equivalent actions (in terms of their consequences) can be performed.

Why are organizations implementing decisions, in their different ways? The need for decision taking arises from different arguments. Here is a brief and non-exhaustive survey of such arguments.
\begin{description}
\item{\em Transparency.} By having important actions arranged as the effect of decisions the how and why of management activity becomes simpler to grasp for external observers.
\item{\em Fraud prevention.} Properly logged decision processes allow external observers and auditors to monitor the  behavior of an organization, and to guarantee that only permissible arguments are used for making choices and for creating decisions.
\item{\em Responsibility sharing.} Very consequential actions can be based on the outcome of group decision processes, with the benefit that individual participants of the decision process need not carry the full weight of the responsibility for the actions.
\item{\em Speed control.} Only by having a well-organized and well-monitored decision process in place an organization can improve its responsiveness for a variety of external requests and events.
\item{\em Preventing tunnel vision.} Small groups of individuals who are operating or managing in a sustained flow of activity run the risk of getting caught in a so-called tunnel vision: one one way ahead can be imagined. Whereas that is true in an ordinary tunnel, it seldom applies to a more open problem area. A well-organized decision process may
prevent the occurrence of leadership tunnel vision.
\end{description}

\subsection{Decision quality}

A most plausible application of decision taking theory is that it leads to agent abilities, or conjectural abilities following \cite{BDV2011b}, which enable the agent to improve the quality of decisions to which it contributes by participating in the decision taking process, or in the decision making process or in the decision  process. Defining decision quality emerges as a major objective in decision theory development.%
\footnote{In \cite{Trull1966} decision quality is distinguished from decision success, indeed quality measures the likelihood of success rather than the success itself. In \cite{BayleyF2008} decision quality is identified with decision process quality, at the exclusion of decision outcome quality. However, decision outcome in \cite{BayleyF2008} corresponds to decision (outcome) effect in the current paper.}

A decision is of a higher quality (compared to another decision) if one or more of the following criteria hold:
\begin{enumerate}
\item The effect of the decision outcome is more likely to correspond to the decision taking agent's intention. In other words the decision outcome is more realistic. 
Unavoidably the operational context in which an agent is active comes into play.  By managing that context in such a way that decision outcomes are more likely to have the intended impact the quality of decisions improves, even if outcomes are formally identical. Indeed quality assessment of decisions involves effects as well as outcomes.
\item The assumptions on which the motivation for a decision outcome is based have been better validated.
\item The expected difference in terms of consequences between taking the decision and not taking any decision at all (about the same theme in the same context etc.) is higher.
\item The decision turns out to be (more) final rather than that it is to be classified as (more) paving the way towards a subsequent decision%
\footnote{This criterion includes a better robustness of the decision outcome against complaints and legal 
objections, in other words, subsequent decision shaking is not plausible.}
\item The decision is closer to the most impacting decision that might have been taken at the same moment of time by the same agent concerning the same theme, with the same intentions in mind.%
\footnote{It seems to be impossible to decouple the quality assessment of a decision from the intentions of the deciding agent. If those intentions are unknown no quality assessment is possible.}
\item The progression of the decision process that has led to the decision is more in accordance with the protocol that must be followed for the particular kind of decision at hand.
\item The decision process leading to the decision has made better use of available resources.
\end{enumerate}

In \cite{Amason1996} one finds the observation that there is a paradox hidden in the quest for decision quality: if a decision process maker plans to involve different specialists in the decision making process, the risk of disagreement increases. In particular if the specialists are invited to think out of the box such disagreements may flourish. That in turn may lead to conflict which subsequently decreases the likelihood that decision outcomes induces the expected effects. From \cite{Amason1996} one may conclude that only if affective disagreement amongst staff members can be avoided, decision quality profits from inviting staff to disagree on the substance of choices that have to be made.

\subsection{Decision making models}
Literature abounds with models of decision making processes. Different models may best fit different circumstances. A survey of models is given in \cite{TarterH1998} where it is also claimed that model selection cannot be performed on the basis of generally agreed rational arguments, rather it is a matter of contingency. A classic model is the garbage can model of \cite{CohenMO1972}. The garbage can model incorporates an architecture of a problem solving model into a model of a decision making life-cycle.

The definition of decision taking presented above has no bias towards any specific decision making model. To begin with the definition leaves open many degrees of freedom for a definition of decision making. More importantly the models 
exist at a higher level of abstraction where the rationale of different steps constituting a decision making progression is qualified.

Besides decision making models there are organizational paradigms in which decision making can play a more or less pronounced role. In \cite{HuberM1986} a paradigm of organizational design is presented which centers around getting adequate decision processes in place. The definition of decision taking is supposed to be independent of organizational paradigms, though it must be sufficiently flexible to deal with  the various decision making workflows that a particular organizational paradigm might prescribe.

\subsection{Levels of decision making and taking}
An agent $a$ which is involved in decision taking is confronted with the question how many of its actions are decisions. If one drives by car to the state agent in order to take a decision about the acquisition of some real estate, then one may be tempted to reserve the phrases decision taking, decision making, and decision process only to decision regarding real estate ownership, or with an equal level of importance. Other activities like the car driving through dense traffic or the preceding selection of a means of transportation take place at a somewhat lower or at least different if not disjoint level of importance. These other activities may involve decision taking as well. In the extreme every action performed by the agent may be considered the consequence of the outcome of some implicit or explicit decision. When, however, a level of abstraction and a theme $t$  has been chosen it becomes possible to distinguish between agent $a$'s actions that are connected with decisions concerning $t$ and other actions, which then are not said to be part of the decision process, even when seen from the perspective of another theme some of those actions are also to be classified as decisions.

In a modular organization decisions are taken and made in different organizational modules concurrently. Encapsulation and abstraction are needed to analyze to what extent decision taking activity and decision making progressions occurring within a certain module can be influenced and observed from different modules. Encapsulation and abstraction are notions which have been given a quite precise meaning in process algebra (see   \cite{BaetenBastenReniers2009}) and it will be a matter of future work to develop these notions with a special focus on
decision taking and decision making.

\subsection{Hierarchy of themes}
Both for an agent and for an organization comprising many agents different themes may be distinguished about which decision can occur. The simplest picture is a hierarchy of importance, where decision processes of a higher importance take place interleaved with process fragments in which decisions of lower importance are taken. Unfortunately this hierarchy is a difficult matter because so-called low level tasks may involve ``go-no go'' 
decisions to be taken which are safety-critical, e.g. whether or not to make an outing in the mountains on a given day or whether or not to make use of a car with a known technical problem, whether or not to be transported by a driver who one does not really trust, and so on.

The better picture may be that at the top level of the hierarchy there are tactical decisions which are potentially
safety-critical though lacking any strategic importance. Below those there may be several independent layers of decision processes for various themes of strategic importance, such as stock market investment, real estate maintenance, job rotation, and family planning. Below that is a vast number of processes involving necessary decisions which lack strategic importance as well 
as any substantial risks, such as holiday timing, choosing a restaurant, buying furniture, and the fulfillment of social obligations. Below that there are processes which involve numerous choices that don't qualify as decisions such as when to do shopping, when to walk the dog, whether or not to halt for some specific traffic light, when to stop filling one's cup of coffee, when to do some house keeping, where to buy a bottle of wine, which telecom provider to use, when to reload one's mobile phone and so on.

\subsection{Hierarchical decision structure inside an organization}
I will assume for the sake of simplicity that the management structure of some organization is like a tree rather than the more usual matrix, where most nodes are individual agents and some nodes consist of groups of agents, that is executives,  managers, or disciplinary oriented employees. A hierarchy in the form of a downwards hanging tree is probably too  simple a model given the matrix organization that many organizations prefer, but it helps to visualize the complexity of decision taking and making. I will assume that each agent occupies a single node in the tree only, in some cases operating as a member of a group of agents. One may rank individuals according to their distance to the top of the tree. The chair operates at level 0, other members of the board operate at level 1, and so on. Halfway the tree one finds the so-called middle management. Their role is the most complex one. The picture given will be simplified in comparison to a real case but it helps form an intuition.

Consider an agent $a$ at level $k>2$. At the level of $a$ some types of decisions are taken, perhaps with the help of agents of at least the same rank. Typically $a$ may be putting into effect a number of threads, by way of multi-threading with a suitable strategic interleaving, one for each of the decisions to which $a$ has a commitment to bring it about by managing its preparation and then enacting it to be taken. For a single prospective  decision $d$, say of type $D$,  agent $a$ can make use of an interface of actions $I_D$, such actions may involve: planning a meeting, asking a colleague for comment, writing a proposed decision outcome for $d$, writing a preparatory proposed decision outcome for $d$, issuing a staff member acting at level $k+1$ or higher to perform one of these tasks, holding a meeting, asking others to review a text and so on.

This leads to $a$ processing a multi-thread for handling a plurality of decision taking tasks. A the same time, however, $a$ may be carrying out actions that may be classified as decision making (though not taking) for decisions to be taken at level $k-1$, this again leading to a multi-thread of tasks for decisions which will be taken by $a$'s immediate superiors (having rank below $k$ by definition). 
In addition $a$ may be dealing with a multi-thread for tasks that play a role in the decision process for decisions which are going to be taken at level $k-2$. 

Working at level $k$ it is feasible for $a$, in principle to produce preparatory work which may serve as an incentive for staff at levels below $k$ (that is higher in the hierarchy in the sense of being closer to the top) to generate objectives that may involve some decision taking at their level. Indeed some cases $a$ can be more influential by proposing ideas for decisions to be taken at higher levels (lower $k$) and therefore by being involved in decision making but not in the corresponding decision taking,  than by deciding about issues that have been determined as belonging to level $k$ or higher (i.e. $l>k$). 

Looking down in the power structure at each level decisions are taken for which $a$ carries some responsibility. Sometimes such decisions are taken in response of questions emanating from $a$ or  from staff members or staff groups with higher authority (that is having lower rank). In other cases these decisions are taken without having been given a trigger from above in advance. That takes place when a predefined workflow is put into effect and if that workflow calls for a decision to be made.

The combination of these multithreads for each deciding agent  is again put in parallel by  means of a strategic interleaving operator. This kind of composition has been analyzed in detail in \cite{BergstraMiddelburg2006}.

\subsection{Conjectural abilities implied by the proposed theory of decision taking}
Following \cite{BDV2011b} awareness of an agent $a$ of a theory of $X$ may lead supporters of that theory (amongst which its author(s)) to attribute conjectural abilities concerning $X$ to $a$. 

The definition of the notion of decision and its surrounding notions can be considered a theory of decision. Whatever its academic merits, awareness of this theory may constitute an addition to an agents competence profile by  way of the acquisition of additional conjectural abilities. Theories of decision from management science often focus on top level decision taking and supporting processes. At top level, however, the question to what extent one is involved in decision taking or making is often not felt as problematic. 

Lower ranking staff members may have more difficulty in assessing their place in an organization's control system. For middle management the analysis of decision may have some attraction, because it may be helpful for assessing one's contribution. Taking the agent $a$ at management level $k>2$ from the top as an example the following conjectural abilities come to mind:
\begin{enumerate}
\item Being able to organize one's participation in range of different decision processes, by classifying this participation as being merely to the decision process, to decision making, to decision taking, or even  to decision shaking, and by viewing these participations as threads in a hierarchical multi-thread under control of $a$ 
by means of strategic interleaving. (See \cite{BergstraMiddelburg2007,BergstraMiddelburg2006}.)
\item Being aware of the variety of protocols that govern decision processes, as well as of the meta decision 
processes that determine these protocols.
\item Being inclined to assess and to forecast and in some cases to influence the quality of decisions made 
by agents at different levels. 
\item Being able to balance the participation to each of a plurality of decision making threads in such a way that the agent's influence is maximized. 
\item Being aware that each participation to a decision process may be influential, in particular roles that are classified as decision making but not as decision taking, and decision process roles that don't qualify as decision making roles.
\item For an agent operating at level $k$ halfway the organization, the so-called middle management roles: being aware that $a$'s participation, by being involved in the full range of decision processes, may well be intrinsically more complex than the participation of staff positioned at tope levels (0, 1, and perhaps 2), the so-called top-management.
\item Being aware that the quality of decision taking cannot be decoupled from the planning of decision taking. The ability to forecast that a certain kind of decision will  turn out to be effective  is an essential decision process design
capability. This ability need not be confined to so-called top management. That foresight  may also appear as a decision making capability or as a decision process making capability. Indeed, staff members at lower levels of the hierarchy may well consciously trigger chains of events from which decisions that will eventually be taken by higher management will emerge, even if they don't participate in either decision making or decision taking, for that particular kind of decision.
\item Understanding in mechanical terms why a middle manager with a clear sense of direction need not be less influential than a top manager without a clear sense of direction. At the same time, given the non-empty sense of direction aggregated at middle management level, understanding  why top management can often do with much less sense of direction than one might expect. They only need to reinforce what pops up in terms of options for decision making and decision taking.
\item Awareness of the fact that a decision it itself an activity which may be in need of algorithmic control. An important part of preparatory work may consist of the development of an instruction sequence, perhaps equipped with conditions phrased in a short-circuit logic, or more generally a proposition algebra (see \cite{BergstraPonse2010}), which needs to be put into effect in real time in order to compute the decision outcome that must to be produced. The act of deciding, that is the decision (what else can it be), then reduces to the 
decision to put an instruction sequence into effect.%
\footnote{At this point the concepts unfortunately become less clear. If an instruction sequence has been determined in advance (as a part of decision making, or even merely as a part of the preceding decision process) in order to compute the decision outcome when a decision, of a particular kind, is taken, then one must distinguish two cases: (a) the instruction sequence is put in to effect by ``manual operation'' by the decision taking agent, in which case the resulting progression can be termed a decision without hesitation, provided it comes to an end, and (b) the instruction sequence is put into effect by automatic means outside the agent's immediate control, in which case one may prefer to refer to the act of putting into effect as an implicit decision, or even as no decision at all.}
\end{enumerate}

\section{Concluding remarks}\label{Conc}
The first conclusion consists of a brief survey of informatics perspectives on decision taking which summarizes the results of the paper concerning the concept of decision making rather than decision taking.

\subsection{Informatics perspectives on decision making}
Besides providing a detailed definition of decision taking and providing a perspective on that from a viewpoint of informaticology the paper implicitly  provides a description of decision making and an informaticological perspective on that theme. Here is a summary of conclusions about decision making that may be inferred from our
discussions above.
\begin{itemize}
\item Description methods for: (decision outcome) typing, design and architecture of decision processes, (decision process) protocols, modeling in time of decision processes. Given more or less formalized descriptions of decision processes simulation, analysis, and verification can be made available in the context of decision making.
\item Protocol support for decision processes, for instance by way of providing workflow models and assistance.
\item Providing a mechanical perspective on decision thread effectuation in terms of  multi-threading with strategic interleaving. Multithreading allows both the concurrent activity of different decision processes as eel as the interleaving of decision processes with other threads of activity.
\item Dedicated terminology for running processes: progression, trace of a thread, concurrent operation, run of a machine, putting an instruction sequence into effect, precondition, postcondition.
\item Formulating real time aspects of decision taking in terms of preparatory design of instruction sequences making use of conditions phrased in short-circuit logic.\footnote{%
This perspective is speculative at this moment, but if on the one hand conditions become more complicated, and on the other hand the wish to limit energy consumption calls for slower processing and even for giving up the guarantees of deterministic output by allowing more impact of noise, then dynamic valuation of boolean expressions (as part of conditions) becomes an unavoidable reality.}
\item Decision making support systems: in practice many decision support systems mainly provide choice support, but all automated support for the decision process  design and decision process control may count as decision making support.
\item Conceptual clarification about the notion of a decision in relation to, promise, imposition, suggestion, request etc.
\end{itemize}

\subsection{Decision taking as a competence from informaticology}
Under the assumption that a decision outcome is a piece of information, decision taking is an act of information production. Often  decision taking merely amounts to the promotion of a preliminary decision outcome, as produced during decision preparation, to a definite status. Decision taking depends on various concepts (competences, abilities) known from informaticology: adapting information status, information classification, managing communication and even broadcasting, managing information visibility, adherence to protocols, and information processing workflow. Further decision taking  is often embedded in a multi-threaded setting. Assuming in addition that choice, however sophisticated in a practical case, plays a supportive role only, decision taking  becomes primarily a competence which to a significant extent belongs to  informaticology. 

\subsection{Options for future work}
Several options for further work can be imagined.\footnote{%
This section has been revised in 2014, having in mind the subsequent work on decision taking that I 
have actually tried to carry  out, however, no reference to such work is made in order to avoid circular (self) referencing.} 
\begin{enumerate}
\item A important  issue is to clarify the relation between decision and promise. 
Some promises are decisions and some decisions are promises. 
According to many authors (e.g. \cite{Atiyah2003}, \cite{Sheinman2011}), 
though not according to Burgess in \cite{Burgess2005,Burgess2007} and several other works, 
a promise effects some obligation. The viewpoint of Burgess has been elaborated in 
ample detail in~\cite{BergstraBurgess2014}, though yet without development of 
a connection with decision taking.

\item Is decision taking a service which can be provided to other agents and which for that reason can be
 outsourced? This matter may be productively investigated as a sequel to the analysis of 
 sourcing and outsourcing as given in~\cite{BDV2011b,BDV2011c}.
 
\item 
Choice from a given menu of alternatives, and real time choice that involves real time development of a 
potential course of action (for which I propose to use the phrase ``action determination''), both require further conceptual analysis, in view of the idea that some, if not most, acts of choice are not decisions. In similar vein
voting must be analyzed in its contras with decision taking.

\item
Buying and selling are commonly assumed to involve decisions from both parties. Working out a case study 
about decisions that play a role in buying and selling will constitute a relevant test of the 
conceptual framework on decision taking that has been developed above.

\item
Having defined decision taking rather meticulously, the question can be posed which activities require decision taking and which activities don't. What are advantages and disadvantages for an agent 
of maintaining a self imposed limitation to an episode of decision (taking) free activity?

\item Looking down in terms of  the hierarchy to a decision process differs from looking up to it. Decision processes by higher management layers seem to make more sense to lower layers in a hierarchy than the other way around. Higher 
management levels may be inclined to view decision processes at lower levels as redundant and standing in the way of the real work. Viewing upwards it is often the case that decisions are considered to be either mistaken or delayed.

These matters can only be considered in detail once appropriate forms of encapsulation (used for restricting remote influence on decision processes), and abstraction (used for restricting the options for obtaining information by a remote agent about a decision processes), have been adequately developed. I expect that the language of process algebra (see \cite{BaetenBastenReniers2009}), in which encapsulation and abstraction pay a central role, 
will be helpful for developing an approach to this matter.

\item By regarding a decision to be an action (that is an act of decision taking) rather than the output of an action 
I have proposed to disambiguate the  language one uses about decisions. This design decision about how to use a language on decisions may be considered a step backward from the perspective of the theory of concept images
of~\cite{TallVinner1981}. Indeed if one first develops a concept image about the notion of a decision one may find contradicting aspects. Now~\cite{GrayTall1994}, carrying on along the same lines, concludes remarkably that persons able to deal in a flexible manner with inconsistent concept images (for instance about limits of sequences and about differentiation of real functions of a single variable) 
in the area of elementary mathematics are better performers than those who in an artificial manner apply refined and consistent concept images for the same notions. 

If a similar conclusion can be drawn about decision taking that will mean that by assimilating our theory of decision taking and decision outcomes (which I will call OODT for outcome oriented decision taking) a person runs the risk of artificially creating a coherent concept image for decision making and as a consequence that person's ability for decision making may be reduced. Obviously it is important to find out such matters before advertising OODT as a practical tool.

\item From a philosophical perspective one may ask if process product ambiguity is an essential feature of the concept of a decision rather than an accidental one. If so then resolving that ambiguity in either way will create larger problems than it solves. Is there some form of tradeoff in this area? Is it possible that for some fields of decision taking the OODT perspective is helpful whereas for other areas a so-called proceptual view in the sense of~\cite{TallVinner1981}
is unavoidable on philosophical grounds?

\item In~\cite{BDV2011b,BDV2011c} process product ambiguity is considered an issue for outsourcing, and
a preference has been formulated for a process interpretation of outsourcing, that is viewing outsoucing as an
action or a process that modifies a so-called sourcement, thereby producing a new sour cement as its outcome.  Many
authors, however, view outsourcing as a specific kind of sour cement, that is a product (of an  outsourcing process
rather than the process itself).
Whether or not attempts at resolving process product ambiguities for
widely used notions are useful at all remains to bee seen. In particular the question may be posed if 
resolving such ambiguities for terms with less ubiquitous use should be tried out. A possible simpler case of an ambiguity is in the notion of a fraction: as a procept a fraction signifies at the same time the result of a division, that is a rational number, and a particular notation for that outcome. Moving from this inconsistent concept image (in the sense of~\cite{TallVinner1981}) to a consistent one (an attempt made for instance in~\cite{Rollnik2009}) may be considered a relevant case study, that might precede an exercise in disambiguating concept 
image refinement for decision taking.
\end{enumerate}

Undeniably last three issues constitute fundamental questions, if not doubts, 
concerning the rationale of my efforts towards conceptual clarification and
corresponding definition of decisions.

\end{document}